\documentstyle[twoside,fleqn,espcrc2,psfig]{article}
\newcommand{\lsim}{\raisebox{0.3mm}{\em $\, <$} 
\hspace{-3.3mm} \raisebox{-1.8mm}{\em $\sim \,$}}

\def\lsim{\raise0.3ex\hbox{$\;<$\kern-0.75em\raise-1.1ex
\hbox{$\sim\;$}}}
\def\gsim{\raise0.3ex\hbox{$\;>$\kern-0.75em\raise-1.1ex
\hbox{$\sim\;$}}}

\begin{document}

\title{CERN to Gran Sasso; An ideal distance for superbeam?
\thanks{Talk presented at 7th International Workshop on Topics in 
Astroparticle and Underground Physics (TAUP2001), 
Laboratori Nazionali del Gran Sasso, Italy, September 8-12, 2001.
}}

\author{Hisakazu Minakata$^a$ and Hiroshi Nunokawa$^{b,c}$ \\
        {\ }\\
$^a$
Department of Physics, Tokyo Metropolitan University, 
1-1 Minami-Osawa, Hachioji \\
Tokyo 192-0397, Japan \\ 
\vglue 0.15cm
$^b$
Instituto de F\'{\i}sica Te\'orica, Universidade Estadual Paulista, 
Rua Pamplona 145 \\ 
01405-900 S\~ao Paulo, SP Brazil \\
\vglue 0.15cm
$^c$
Instituto de F\'{\i}sica Gleb Wataghin, 
Universidade Estadual de Campinas -- UNICAMP\\
P.O. Box 6165, 13083-970 Campinas SP Brazil
} 

\begin{abstract}
We use the CP trajectory diagram as a tool for pictorial 
representation of the genuine CP and the matter effects to 
explore the possibility of an {\it in situ} simultaneous 
measurement of $\delta$ and the sign of $\Delta m^2_{13}$. 
We end up with a low-energy conventional superbeam experiment 
with a megaton-class water Cherenkov detector and baseline length 
of about 700 km. 
A picturesque description of the combined ambiguity which may 
arise in simultaneous determination of $\theta_{13}$ and 
the above two quantities is given in terms of CP trajectory diagram. 

\end{abstract}

\maketitle

%%%%%%%%%%%%%%%%%%%%%%%%%%%%%%%%%%%%%%%%%%%%%%%%%%%%%%%%%%%%%%%%%%%%%%
%% Section I %%%%%%%%%%%%%%%%%%%%%%%%%%%%%%%%%%%%%%%%%%%%%%%%%%%%%%%%%
%%%%%%%%%%%%%%%%%%%%%%%%%%%%%%%%%%%%%%%%%%%%%%%%%%%%%%%%%%%%%%%%%%%%%%

\newpage
%%%%%%%%%%%%%%%%%%%%%%%%
\section{Introduction}

Exploring the structure of lepton flavor mixing and the neutrino 
mass pattern is one of the most challenging goals of contemporary 
particle physics. Among other things, the least known is the 
(1-3) sector of the MNS matrix, the angle $\theta_{13}$, the sign of 
$\Delta m^2_{13}$, and the CP violating angle $\delta$. 
(See e.g., ref. \cite {NOW2000mina}.)

We have explored in a series of papers \cite {MN} the features of 
interplay between the CP phase and matter effect, and tried to 
develop strategies for measurement of lepton CP violation in 
long-baseline neutrino oscillation experiments. 
See also ref. \cite {cp-matter} for related works.
Along the line of thought, we have introduced recently a 
powerful tool called ``CP trajectory diagram in bi-probability space'' 
which allows us a separate pictorial representation of the genuine 
CP and the matter effects \cite {MNjhep01}. 
We pointed out that an ambiguity exists in determination 
of these parameters in a correlated way 
($\delta$ $-$ sign of $\Delta m^2_{13}$). 

We address here the issue of an {\it in situ} simultaneous 
determination of $\delta$ and the sign of $\Delta m^2_{13}$ 
in a single experiment. 
In a companion article \cite {nufact01}, which is a 
contribution to Proceedings of NuFACT01, a concise summary 
of the idea of CP trajectory diagram and its use is given and the 
principle of optimizing beam energy is discussed. 

Let us start with a brief remark on the more generic feature 
of the ambiguity problem. 

\section{Clover-leaf ambiguity}

If the value of $\theta_{13}$ is unknown, there exists another 
ambiguity in a correlation ($\delta - \theta_{13}$), 
as pointed out in ref. \cite {B-Castell}.
Together with the ambiguity we have uncovered, there exists the  
combined ambiguity which can be as large as four-fold. 
We now demonstrate that it can be described 
in a simple picturesque way by using the CP trajectory diagram.

\begin{figure}[ht]
\vglue -0.2cm 
\centerline{\protect\hbox{
\psfig{file=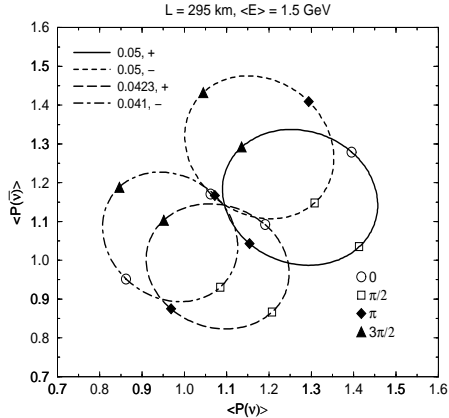,height=6.2cm,width=6.8cm}
}}
\vglue -1.4cm 
\caption{\small
Illustration of the clover-leaf ambiguity in termes of CP 
trajectory diagram; four solutions exist for given values of 
$P(\nu_\mu\to\nu_e)$ and $P(\bar{\nu}_\mu\to\bar{\nu}_e)$ 
at neutrino energy $\langle E \rangle=1.5$ 
GeV and baseline distance $L=295$ km. 
Corresponding values of $\sin^22\theta_{13}$ as well as 
the sign of $\Delta m^2_{13}$ whose absolute value is 
$3 \times 10^{-3}$ eV$^2$ are indicated in the plot.  
The remaining mixing parameters are taken as;
$\sin^22\theta_{23} = 1.0$, 
$\Delta m^2_{12} = 5 \times 10^{-5}$ eV$^2$, 
$\sin^22\theta_{12} = 0.8$, and 
$\delta = \pi/2$. 
We take the matter density as $\rho = 2.72$ g/cm$^3$ and 
the electron fraction as $Y_e$ = 0.5. 
}
\label{Fig1}
\vglue -0.5cm 
\end{figure}

In fig. \ref{Fig1} drawn is the CP trajectory diagram for 
four values of the mixing parameters which are given in the 
caption of fig. \ref{Fig1} assuming, for simplicity, 
neutrino beam with the Gaussian type energy distribution
as used in our recent work \cite {MNjhep01}. 
The point of fig. \ref{Fig1} is that the four-fold solutions 
are possible for given oscillation probabilities of 
$P(\nu)\equiv P(\nu_{\mu} \rightarrow \nu_{e})$ and 
$P(\bar{\nu}) \equiv P(\bar{\nu}_{\mu} \rightarrow \bar{\nu}_{e})$, 
both at about 1.1 \% in fig. \ref{Fig1}.
One notices from fig. \ref{Fig1} that the name 
"clover-leaf ambiguity" by which the ambiguity is referred at 
TAUP2001 is quite natural and appealing.

%%%%%%%%%%%%%%%%%%%%%%%%%%%%%%%%%%%%%%%%%%%%%%%%%%%%%%%%%%
%%%%%%%%%%%%%%%%%%%%%%%%%%%%%%%%%%%%%%%%%%%%%%%%%%%%%%%%%%%%%%%

\section {Optimal distance for measuring the sign of $\Delta m^2_{13}$}

Now we turn to the discussion of our original question, i.e., 
how to resolve the ($\delta$ $-$ sign of $\Delta m^2_{13}$) ambiguity. 
Since the sign of $\Delta m^2_{13}$ can be determined by measuring 
interference between the vacuum and the matter effects, it is 
natural to think about neutrino oscillation experiments which 
utilize longer baselines. One can think of these possibilities in 
the context of either  

\noindent
(i) a single detector experiment for {\it in situ} simultaneous 
determination of $\delta$ and the sign of $\Delta m^2_{13}$, or

\noindent
(ii) a two-detector experiment with second supplemental detector 
which is primarily devoted for determination of the sign of 
$\Delta m^2_{13}$.

The next question to ask is; what is the optimal baseline length 
for this purpose? It would be the best situation if we can tune the 
distance in such a way that the matter effect is relatively enhanced 
compared to the genuine CP violating effect. 
To quantify the request we define the asymmetry parameter defined 
by using the ratio 
$R(P) \equiv \langle P(\nu_\mu \to \nu_e) \rangle /
\langle P(\bar{\nu}_\mu \to \bar{\nu}_e) \rangle$ as 
\begin{equation}
A(R) \equiv 
\frac{R(P;\Delta m_{13}^2>0)-R(P;\Delta m_{13}^2<0)}
{R(P;\Delta m_{13}^2>0)+R(P;\Delta m_{13}^2<0)},
\label{asymmetry}
\end{equation}
where the probabilities are averaged over Gaussian
type energy distributions. 
By using the ratio in defining the asymmetry it is 
insensitive to the values of the mixing parameters, 
in particular to $\delta$.

\begin{figure}[ht]
\vglue -0.9cm 
\centerline{\protect\hbox{\hglue -1.0cm 
\psfig{file=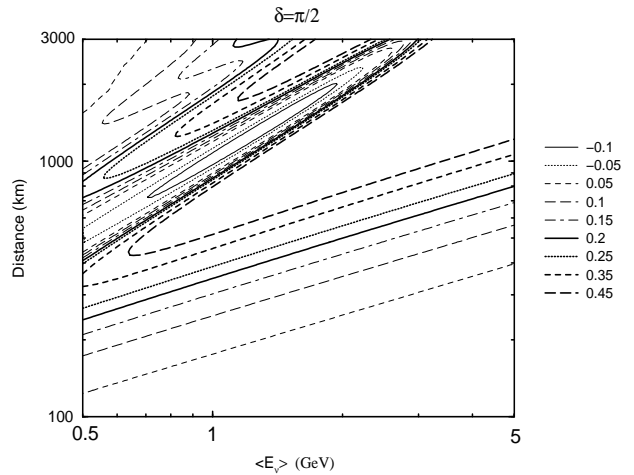,height=6.7cm,width=7.7cm}
}}
\vglue -1.3cm 
\caption{\small
Asymmetry of the probability ratio defined in Eq. (\ref{asymmetry})
in the text computed for $\delta = \pi/2$.
The mixing parameters are chosen as the same with those of 
fig. \ref{Fig1} and $\sin^22\theta_{13} = 0.05$.
}
\label{Fig2}
\vglue -1.0cm 
\end{figure}

In fig. \ref{Fig2}, the asymmetry $A(R)$ is plotted for 
$\delta = \pi/2$ with the same mixing parameters as in 
fig. \ref{Fig1}, apart from fixing $\sin^22\theta_{13}$ 
to be 0.05. One notices that the asymmetry 
is large at $L = 600-700$ km, and at $1000-1500$ km 
for $E \sim 1$ GeV. It can be explicitly shown that 
the asymmetry is indeed very insensitive to $\delta$ 
\cite {MNjhep01}. Therefore, the baselines $L \sim 700$ km, 
and at $\sim 1000$ km look promising. We take the former 
option and examine the feature of CP-matter interplay by 
using the CP trajectory diagram.

%%%%%%%%%%%%%%%%%%%%%%%%%%%%%%%%%%%%%%%%%%%%%%%%%%%%%%%%%%
%%%%%%%%%%%%%%%%%%%%%%%%%%%%%%%%%%%%%%%%%%%%%%%%%%%%%%%%%%

\vglue -0.5cm 
\section {Longer baseline option; single vs. two-detector methods}

\begin{figure}[ht]
\vglue -0.5cm 
\centerline{\protect\hbox{
\psfig{file=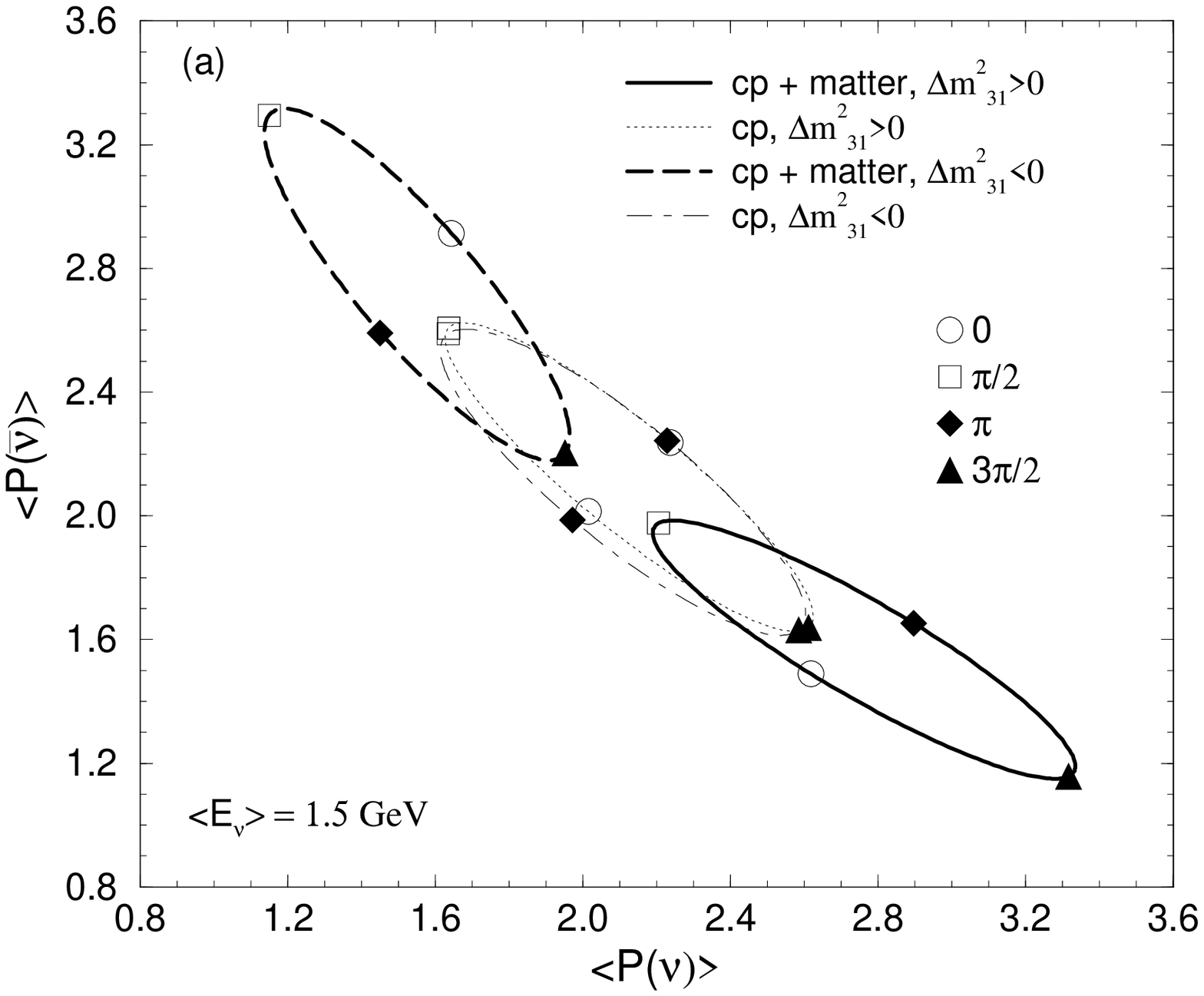,height=7.0cm,width=7.6cm}
}}
\vglue -1.0cm 
\centerline{\protect\hbox{
\psfig{file=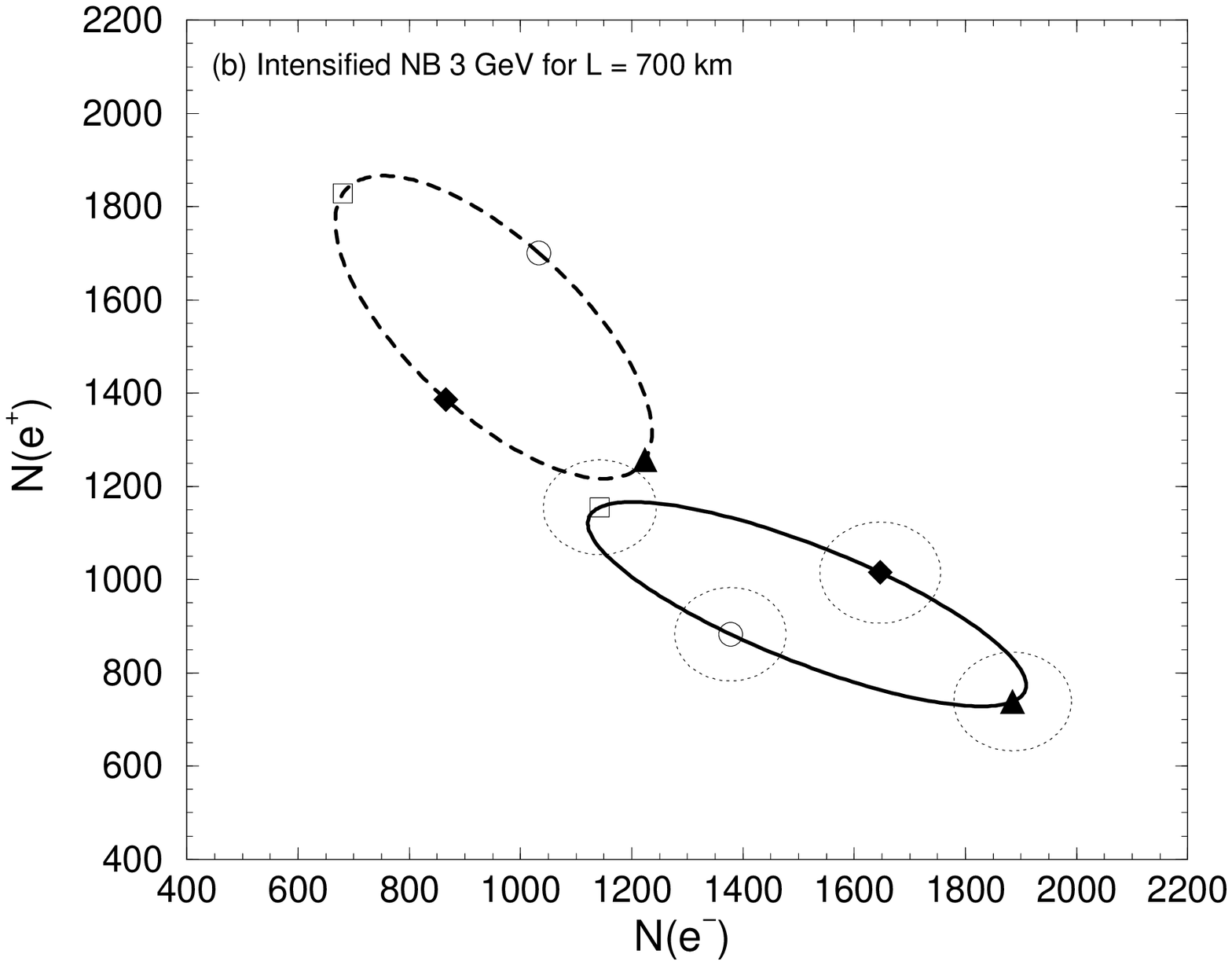,height=7.0cm,width=7.6cm}
}}
\vglue -1.5cm 
\caption{\small
CP trajectory diagrams in 
(a) bi-probability space, and in 
(b) event number space $N(e^-) - N(e^+)$ 
for baseline distance of $L=700$ km.
In (b) a factor of 3 intensified NB 3 GeV beam is used to imitate
OA beam of peak energy 1.4 GeV, and the dotted circles 
correspond to 3 $\sigma$ statistical uncertainty.}
\label{Fig3}
\vglue -0.6cm 
\end{figure}

In fig. \ref{Fig3}, we present (a) CP trajectory diagram 
in bi-probability plane for neutrino beam with the Gaussian type 
energy distribution with averaged energy $\langle E \rangle =1.5$ GeV 
and baseline distance $L=700$ km, and 
(b) CP trajectory diagram on number of events plane with the 
same baseline for appearance channels 
$\nu_{\mu} \rightarrow \nu_{e}$ and 
$\bar{\nu}_{\mu} \rightarrow \bar{\nu}_{e}$.  

One can clearly see in fig. \ref{Fig3}a that the two 
trajectories corresponding to 
$\Delta m^2_{13} >0$ (solid line) and 
$\Delta m^2_{13} <0$ (dashed line) 
are well separated with each other. Therefore, it is 
in principle possible to carry out an {\it in situ}
simultaneous measurement of $\delta$ and the sign of 
$\Delta m^2_{13}$ in such a baseline length. 
It is amusing to note that the length just corresponds to 
either CERN $\rightarrow$ Gran Sasso, or 
Fermilab $\rightarrow$ Soudan mine distances.

In fig. \ref{Fig3}b, we assume a water Cherenkov detector of 
fiducial volume 0.9 Mton, and 4 MW of proton beam power 
which is planned in the JHF experiment in its phase II \cite {JHF}.
We use the narrow band (NB) 3 GeV beam whose neutrino energy peaks 
at $E \sim 1.4$ GeV considered in ref. \cite {JHF}, 
but with intensity multiplied by factor of 3. It is to 
mimic the off-axis (OA) beam which is designed for  $E \sim 1.4$ GeV, 
a bit of higher energy than the one actually prepared for the 
JHF experiment\cite {JHF}. 
Two (six) years of running is assumed for neutrino 
(antineutrino) channel.
We refer ref. \cite {MNjhep01} for a detailed explanation of how the 
computation of number of events is done.  

As you see in fig. \ref{Fig3}b, the numbers of events are sizable, 
some 1000 $-$ 2000, though not gigantic. Resultant 3 $\sigma$ 
contour is small enough so that simultaneous measurement of 
$\delta$ and the sign of $\Delta m^2_{13}$ seems feasible, 
justifying the title of this talk.  

If the detector at baseline $L=700$ km cannot be so massive 
by some reasons, or if the practical site requires much longer 
distance, then it can be viewed as a secondary (farthest) detector, 
assuming that the primary far detector (e.g., Hyper-Kamiokande 
\cite {totsuka} in the case of the JHF experiment) already exists. 
In this case, the secondary detector is primarily for measurement 
of the sign of $\Delta m^2_{13}$ and the requirement of statistics 
can be relaxed.

Finally, it was pointed out in the talk that while the present tunnel 
in the Gran Sasso Laboratory is too small to accommodate a megaton water 
Cherenkov detector, a 180 kton "proto-type" is ready to be 
created upon filling water into the whole tunnel!

%%%%%%%%%%%%%%%%%%%%%%%%%%%%%%%%%%%%%%%%%%%%%%%%%%%%%%%%%%
%%%%%%%%%%%%%%%%%%%%%%%%%%%%%%%%%%%%%%%%%%%%%%%%%%%%%%%%%%

%%%%%%%%%%%%%%%%
\section*{Acknowledgments}
%%%%%%%%%%%%%%%%

This work was supported by the Brazilian funding agency
%Funda\c{c}\~ao de Amparo \`a Pesquisa do Estado de S\~ao Paulo (FAPESP),
FAPESP, 
and by the Grant-in-Aid for Scientific Research in Priority Areas
No. 12047222, Japan Ministry of Education, Culture, Sports, Science
and Technology.

%%%%%%%%%%%%%%%%%%%%%%%%%% Bibliography %%%%%%%%%%%%%%%%%%%%%%%%%%%%%%

\end{document}